\documentclass[3p,times,procedia]{elsarticle}
\usepackage{nupha_ecrc}

\volume{00}

\firstpage{1}

\journalname{Nuclear Physics A}

\runauth{}

\jid{nupha}

\jnltitlelogo{Nuclear Physics A}

\usepackage{amssymb}
\usepackage[figuresright]{rotating}
\usepackage{amsmath}
\usepackage{subcaption}

\begin{document}

\begin{frontmatter}



%% Instructions from Editor: Please use the following \dochead only in the preprint version (e-print arXiv etc.); 
%% use empty \dochead{} when submitting to Nuclear Physics A!
\dochead{XXVIIth International Conference on Ultrarelativistic Nucleus-Nucleus Collisions\\ (Quark Matter 2018)}
%% e.g. \dochead{17th International Conference on Dynamical Processes in Excited States of Solids}

\title{Jet modification with medium recoil in quark-gluon plasma}


\author{Chanwook Park}
\author{Sangyong Jeon}
\author{Charles Gale}

\address{Department of Physics, McGIll University, 3600 University Street, Montreal, QC, H3A 2T8, Canada}

\begin{abstract}
Jet energy transported to quark-gluon plasma during jet-medium interaction excites the QGP medium and creates energetic thermal partons -- recoil particles or recoils.
Modification of the jet structure in heavy ion collisions is studied using \textsc{martini}, in which recoil simulation is enabled.
In large systems such as central Pb-Pb collisions, the recoil effect is expected to be critical due to strong jet-medium interaction.
We show the results of the jet mass function and jet shape function are improved when the recoil particles are included in the reconstructed jets.
We conclude that the energy carried by the recoil particles are regarded as a part of reconstructed jets and are necessary in studying jet modification in heavy ion collisions.

\end{abstract}

\begin{keyword}
jet energy loss \sep recoil particles \sep full jet reconstruction


\end{keyword}

\end{frontmatter}

% \linenumbers


\section{Introduction}
\label{}

In heavy ion collisions, jets and thermal medium continuously interact each other; the jets are quenched due to the medium while inducing medium excitation along the jet path.
The energy lost by jets, present as a form of medium response or recoil, gets dissipated by the flow of the medium, but some of the energy may still remain in the jet cone.
When defining jets using the full jet reconstruction techniques, one should take into account whole energy-momentum within a given jet cone only excluding those originated from the thermal medium.

Using \textsc{martini}~\cite{Schenke:2009gb}, in which recoil simulation is newly implemented, we present the effect of recoils in studying jet structure observables, i.e., jet invariant mass and the jet shape function.
We find that the jet mass is reduced by jet quenching, while recoils substantially enhance the jet mass especially for higher energy jets.
Also we show that recoils are crucial in describing the jet shape function at the peripheral side of a jet cone, giving rise to a increasing trend of the ratio between PbPb and pp collisions up to $\Delta r = 1$.

\section{MARTINI}

% General MARTINI description
\textsc{martini} is a Monte Carlo event generator for jet evolution in high-energy heavy ion collisions~\cite{Schenke:2009gb}.
It performs in-medium parton shower according to the AMY formalism for the radiative energy loss rates~\cite{Arnold:2002ja,Arnold:2002zm} combined with collisional processes~\cite{Schenke:2009ik}.
The energy loss rates depend on the properties of QGP medium, which is provided by hydrodynamics simulations.
The initial vacuum shower and hadronization of the evolved partons are accomplished by \textsc{pythia 8}~\cite{Sjostrand:2014zea}.

The time-evolving energy distribution $P_a(p, t)$ for a given particle $a$ can be expressed in terms of a set of coupled rate equations, which takes the following form~\cite{Schenke:2009gb} :

\begin{align} 
	\frac{dP_q(p)}{dt} = \int\limits_k & P_q(p+k)\frac{d\Gamma^{q}_{qg}(p+k,k)}{dkdt}
	- P_q(p)\frac{d\Gamma^{q}_{qg}(p,k)}{dkdt}
	+2P_g(p+k)\frac{d\Gamma^g_{q\bar{q}}(p+k,k)}{dkdt},\\
	\frac{dP_g(p)}{dt} = \int\limits_k & P_q(p+k)\frac{d\Gamma^{q}_{qg}(p+k,p)}{dkdt}
	+ P_g(p+k)\frac{d\Gamma^{g}_{gg}(p+k,p)}{dkdt}
	-P_g(p)\left(\frac{d\Gamma^g_{q\bar{q}}(p,k)}{dkdt} + \frac{d\Gamma^g_{gg}(p,k)}{dkdt}\theta(2k-p)\right).
\end{align}

$d\Gamma^a_{bc}(p,k)/dkdt$ is the transition rate for a process where a parton $a$ of energy $p$ emits a parton $c$ of energy $k$ and becomes a parton $b$.

The AMY formalism describes energy loss of hard jets in heavy ion collisions as parton bremsstrahlung in the evolving QGP medium. 
The effective kinetic theory described in~\cite{Arnold:2002zm} assumes that quarks and gluons in the medium are well defined (hard) quasi-particles, whose typical momentum is $p \gg T$ and thermal mass is order $gT$.
Under this assumption, the radiation rate can be calculated by means of integral equations~\cite{Arnold:2002ja}.
Radiation is strictly collinear at the splitting vertex while collisional processes involve space-like momentum transfer inducing jet momentum broadening.

% Finite-size effect and running coupling
The radiative energy loss mechanism is improved by implementing the effects of finite formation time~\cite{CaronHuot:2010bp} and running coupling~\cite{Young:2012dv}.
The formation time of the radiation process increases with $p/p_T^2$, and a hard parton and an emitted parton are coherent within that time.
This interference effect suppresses the radiation rate at early times after the original radiation.
For the renormalization scale of running coupling constant $\alpha_s(\mu)$, we use the root mean square of the momentum transfer $\sqrt{\langle p^2_\perp \rangle}$ between the two particles, parameterized as

\begin{equation}
	\sqrt{\langle p^2_\perp \rangle} = (\hat{q}p)^{1/4},
\end{equation}
where $\hat{q}$ is the averaged momentum transfer squared per scattering and $p$ the energy of the mother parton.

\begin{figure*}[t]
	\centering
	\includegraphics[width=0.45\textwidth]{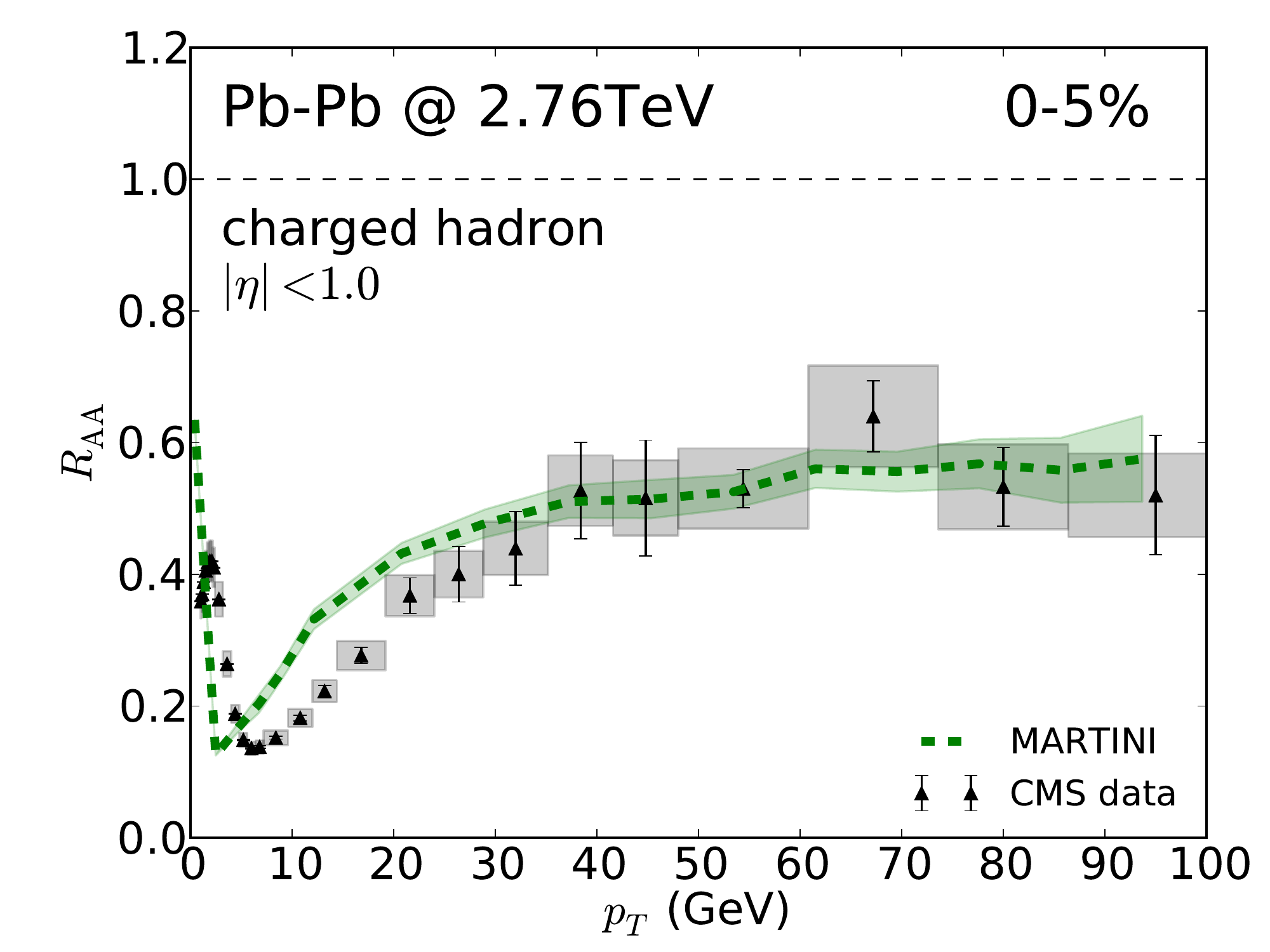}
	\caption[]{The nuclear modification factor $R_{\rm AA}$ of charged hadron in 0-5\% Pb-Pb collisions at $\sqrt{s}$ = 2.76TeV, compared with the CMS measurement~\cite{CMS:2012aa}.}
	\label{fig:RAA}
\end{figure*}

% Implementation of recoil
A recoil parton is generated by adding momentum transfer and momentum of a thermal parton sampled from the medium. 
If the summed momentum satisfying the on-shell condition is greater than a certain kinematic cut, $p_{cut}$, this is promoted to a recoil parton and further can participate in jet-medium interactions.
In this work, we set $p_{cut} \simeq 4T$, where $T$ is the temperature of thermal medium.
Any momentum contributions softer than $p_{cut}$ should be treated as sources for medium response, which is our future work.

% Setup for this work and charged hadron RAA
The thermal background is produced by \textsc{music}~\cite{Schenke:2010nt}, which allows full 3+1 dimensional hydrodynamics calculations.
Thermal fluctuation in the transverse plane is initialized by the IP-Glasma model~\cite{McDonald:2016vlt}.
As shown in Fig.~\ref{fig:RAA}, the \textsc{martini} calculation of the nuclear modification factor $R_{AA}$ is consistent with the CMS data~\cite{CMS:2012aa}, especially at high $p_T$ region.
This indicates that \textsc{martini} is valid for describing leading-order jet fragmentation.

\section{Jet mass}
Unlike leading hadron observables, such as $R_{AA}$, jet structure observables take into account  the distribution of energy-momentum inside a reconstructed jet cone. 
As a part of jet constituents, recoils contribute to the total jet energy and the jet energy distribution with respect to the jet axis, therefore they play an essential role in the modification of jet structures.

\begin{figure*}[t]
	\centering
	\begin{subfigure}[b]{0.7\textwidth}
		\centering
		\includegraphics[width=\textwidth]{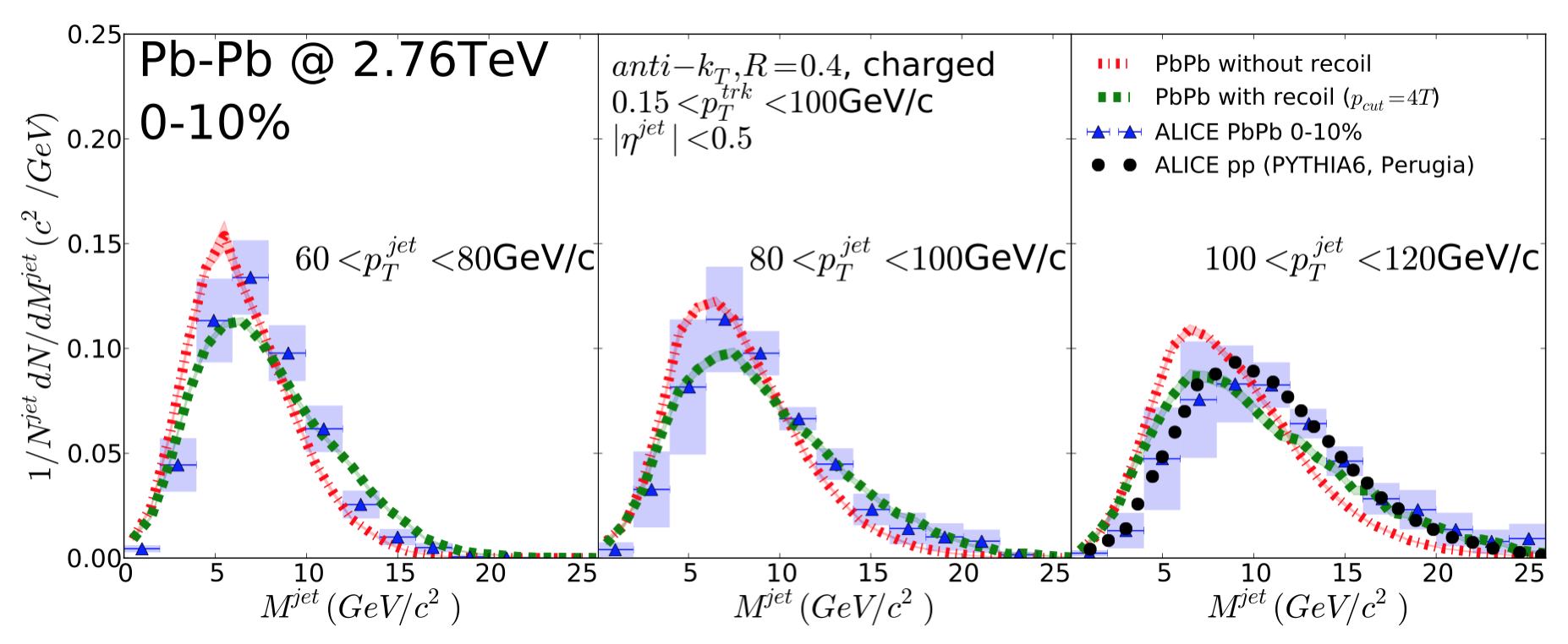}
	\end{subfigure}
	\begin{subfigure}[b]{0.28\textwidth}
		\centering
		\includegraphics[width=\textwidth]{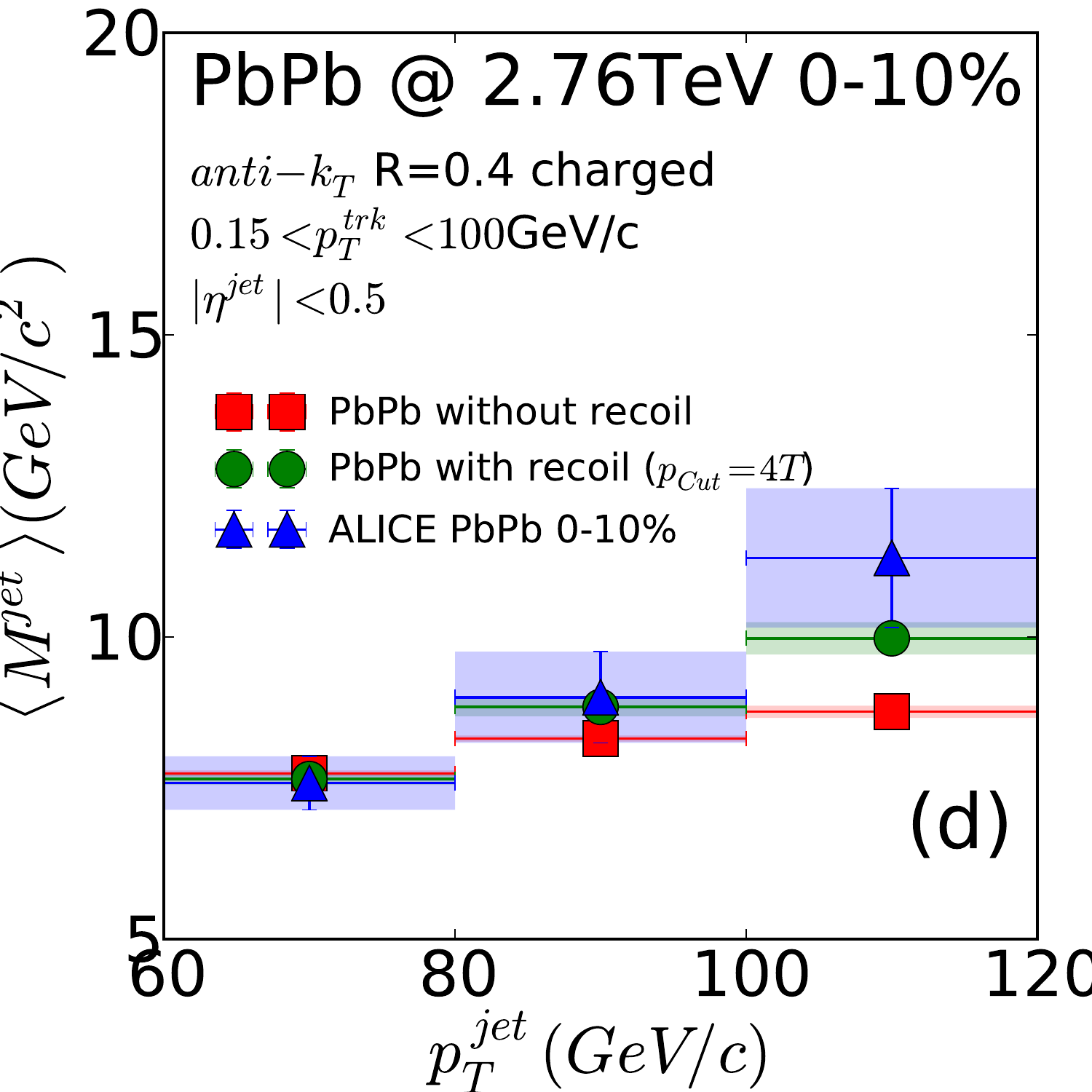}
	\end{subfigure}
	\caption[]{(a-c) The jet mass distribution in $0$-$10\%$ Pb-Pb collisions at $2.76$TeV for 3 different jet $p_T$ intervals. The pp (black) result from ALICE is shown in the right panel for the comparison purpose. 
	(d) The averaged jet mass as a function of jet $p_T$. The measurements are taken from ALICE experiments~\cite{Acharya:2017goa}}
	\label{fig:jet_mass}
\end{figure*}

Jet mass can be changed by jet quenching and broadening in thermal medium. Fig.~\ref{fig:jet_mass} (a) shows the normalized jet mass distribution function for 3 different jet $p_T$ windows in central Pb-Pb collisions at $2.76$TeV.
The jet mass distributions for all jet $p_T$ ranges are altered toward higher jet mass in the presence of recoils.
As shown in the right panel, the measurement for pp and Pb-Pb collisions are fairly coincident.
Two effects, jet quenching induced by QGP and the creation of recoils, have competing influences and the recoils help restoring the jet mass reduced by jet quenching.
The effect of increasing jet mass can be clearly seen in the averaged jet mass plot, shown in Fig.~\ref{fig:jet_mass} (b).
The contribution of recoils to the average of jet mass is greater for higher jet $p_T$ intervals and becomes more important to describe the data measured by ALICE experiment~\cite{Acharya:2017goa}.

\section{Jet shape function}

\begin{figure*}[t]
	\centering
	\begin{subfigure}[b]{0.4\textwidth}
		\centering
		\includegraphics[width=\textwidth]{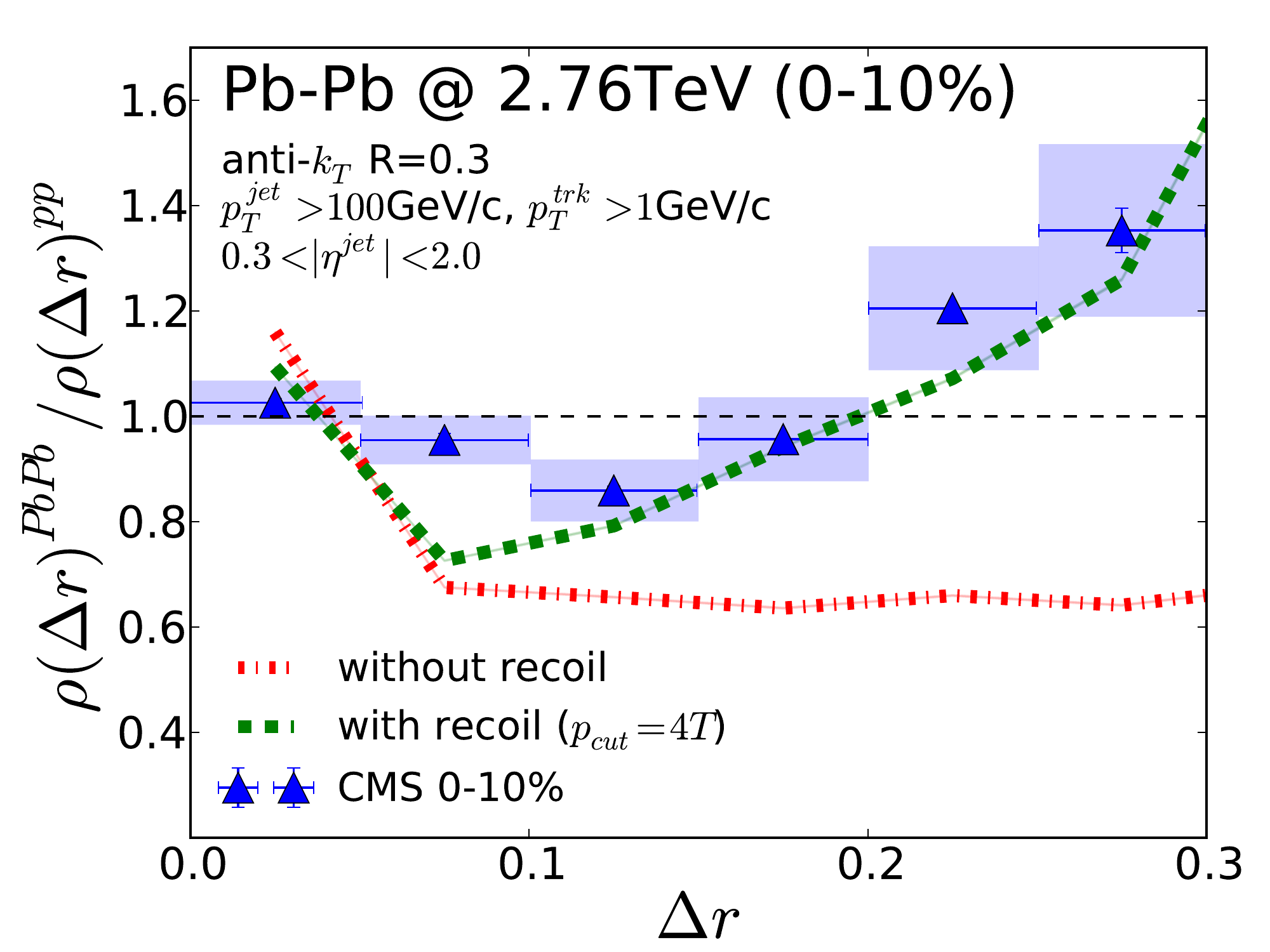}
	\end{subfigure}
	\begin{subfigure}[b]{0.4\textwidth}
		\centering
		\includegraphics[width=\textwidth]{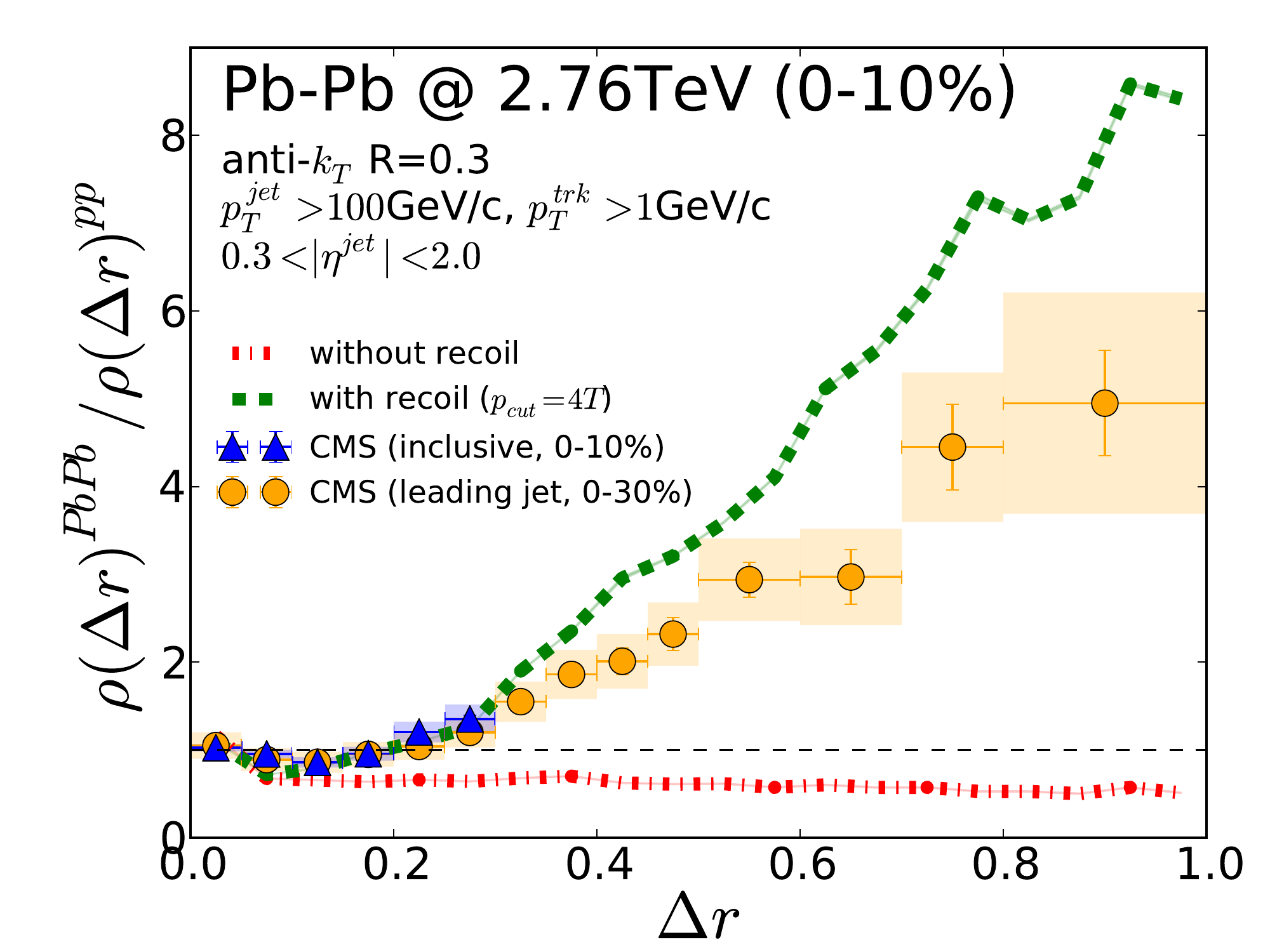}
	\end{subfigure}
	\caption[]{(a) The ratio of the jet shape function between $0-10\%$ PbPb and pp collisions at 2.76TeV. (b): Same as (a) except that $\Delta r$ is extended to $1$. The CMS measurements are compared~\cite{Chatrchyan:2013kwa,Khachatryan:2016tfj}.}
	\label{fig:jet_shape_function}
\end{figure*}

Fig~\ref{fig:jet_shape_function} (a) shows the ratio of the jet shape function in central PbPb collisions at $2.76$TeV and in pp, compared to CMS measurement~\cite{Chatrchyan:2013kwa}.
The \textsc{martini} calculation excluding recoils results in narrower jets with monotonic $\Delta r$ dependence of the ratio at large angles due to jet quenching.
Meanwhile recoils greatly affect the jet shape function at larger angles and give rise to an increase in the ratio of the functions.
We obtain a good agreement between the \textsc{martini} simulation including recoils and the experimental measurement.

In Fig.~\ref{fig:jet_shape_function} (b), the same plot, but the $\Delta r$ extended to $1$, is shown.
The CMS measurement for leading jets in the $0$-$30\%$ bin~\cite{Khachatryan:2016tfj} is also shown for a rough comparison.
Without recoils, jets are consistently quenched in a broad area outside of the jet cone, while the influence of recoils gets bigger at larger $\Delta r$.
We find a rising trend of the ratio showing reasonable slope.
This indicates that our simulations with recoils give an adequate description for the jet shape function of quenched jet in the hydrodynamic medium.

\section{Conclusion}

In this proceedings, we present the importance of recoils in studying jet modification in heavy ion collisions using \textsc{martini} in which the recoil simulation is implemented.
With this prescription, we successfully reproduced the jet mass function and jet shape function by including recoils.
We found that the recoils greatly contribute to the jet mass function for higher $p_T$ jets and the ratio of the jet shape functions especially at peripheral side of the jets.
Our results indicate that jet energy transferred to thermal medium must be counted as a part of the jet substructure. 
This study manifests that recoils are necessary when defining jets quenched by the QGP in heavy ion collisions. 

\section{Acknowledgements}

This work was supported in part by the Natural Sciences and Engineering Research Council of Canada. 
C.G. gratefully acknowledges the Canada Council for the Arts for funding through its Killam Research Fellowship Program. 
Computation for this work was done in part on the Cedar cluster maintained by WestGrid and Compute Canada and on Guillimin cluster managed by Calcul Queb\'ec and Compute Canada.



%% References


\bibliographystyle{elsarticle-num}
\bibliography{CPark}






% \end{thebibliography}

\end{document}